\documentstyle[aps,multicol,epsf]{revtex}

\begin{document}
\draft
\title{Scaling behavior of developing and decaying networks}

\author{S.N. Dorogovtsev$^{1, 2, \ast}$ and J.F.F. Mendes$^{1,\dagger}$}

\address{
$^{1}$ Departamento de F\'\i sica and Centro de F\'\i sica do Porto, Faculdade de Ci\^encias, 
Universidade do Porto\\
Rua do Campo Alegre 687, 4169-007 Porto, Portugal\\
$^{2}$ A.F. Ioffe Physico-Technical Institute, 194021 St. Petersburg, Russia 
}

\maketitle

\begin{abstract}
We find that a wide class of developing and decaying networks has scaling properties similar to those that were recently observed by Barab\'{a}si and Albert in the particular case of growing networks. 
The networks considered here evolve according to the following rules: 
(i) Each instant a new site is added, the probability of its connection to old sites is proportional to their connectivities. 
(ii) In addition, 
(a) new links between some old sites appear with probability proportional to the product of their connectivities or (b) some links between old sites are removed with equal probability. 
\end{abstract}





Recently observed scaling behaviour of a number of networks 
$[1-14]$ again sharply increased the interest of these exciting objects widely studied for a long time 
$[15-20]$. In fact the scaling properties were found only in a few of a great number of growing networks \cite{r98,ajb99,ba99,asbs00,dmj00} but one of them -- Web -- is so significant for everybody that the topic turned to be really hot. 

The simplest model of a scale-free growing network was proposed by Barab\'{a}si and Albert \cite{ba99,baj99}. In this model, each new site is connected with some old site with probability proportional to its connectivity $k$, i.e. to the number of connections with this site. Then the distribution of the connectivities in the large network has a power-law dependence $P(k) \propto k^{-\gamma}$ with the exponent $\gamma=3$ 
\cite{ba99,baj99}. 
In fact, such a network is self-organized into a scale-free structure. 
A full form of the distribution of the connectivities and some other related properties of the model were calculated exactly \cite{dms00}. 
Introduction of aging of sites 
proportional to $\tau^{-\alpha}$, where $\tau$ is the age of a site, does not change scaling properties crucially for $\alpha < 1$, but scaling breaks at higher values of the aging exponent \cite{dm00}.
Several examples of real networks with aging of sites are described in \cite{asbs00}. The simplicity of the Barab\'{a}si--Albert's (BA) model makes it a convenient object to study evolution of networks.

Nevertheless, the BA model describes only a particular type of evolving networks. Of course, reality is much reacher.
In real networks, e.g. in Internet, links are not only added but may break from time to time. That certainly changes the structure of such networks.  
Note also, that new links between old sites may appear in a different way than links between new and old sites, according to different rules. 
Therefore, it is tempting to find out whether the observed behaviour is usual for a vast variety of networks or applies only to a very restricted number of invented objects. With that purpose, we extend a set of models of evolving networks starting from the BA model.

In the present letter, we propose models of developing and decaying networks with undirected links which show scaling behaviour. We consider structures which evolve due to the following reasons. First, they grow like in the BA model, i.e. in each instant one new site is added 
and is connected with an old site by an undirected link with a probability proportional to its connectivity $k$ (one may check that the general results -- the existence of the scaling and the values of the scaling exponents -- do not depend on the number of the connections with a new site).  
In addition, we introduce a new parallel component of the evolution -- the permanent addition of new undirected links between old sites or, on the contrary, the permanent removal of some old links. We consider two different cases. (a) A developing network: Each instant, new $c$ links are added between unconnected pairs of old sites $i$ and $j$ with probability proportional to the product of their connectivities $k_i k_j$. ($c$ may be also non integer. For that one can introduce probability of addition of a link.) $c \geq 0$. (b) A decaying structure (in fact, it is a set of clusters): Each instant, some links between old sites are removed with equal probability. In this case $c \leq 0$.

Note that both processes -- the addition of new sites with new links and the addition of new links between old sites (or removing old links) proceed in parallel, so the resulting structures differ from the original BA model all the time.

We study the following one-site quantities of the structures: the total distribution of connectivities 
at long times, $P(k)$, and the average connectivity of a site of an age $s$ at long time $t$, $\overline{k}(s,t)$, and their scaling exponents $P(k) \propto k^{-\gamma}$ and $\overline{k}(s,t) \propto (s/t)^{-\beta}$.  

Below, we demonstrate both analytically and by simulation that the introduced evolving networks show scaling behaviour in a wide range of values of $c$. Nevertheless, while both $P(k)$ and $\overline{k}(s,t)$, for the developing networks, are power-law functions for all $c \geq 0$, only $\overline{k}(s,t)$ demonstrates the power-law behaviour 
in the whole range $-1<c<0$ for the decaying structures. In this case, the power-law dependence of the distribution $P(k)$ is observed only close to $c=0$.       

In order to study scaling properties of the evolving networks we performed numerical simulations according to the introduced above rules. Each instant, we add one new site with one link and, in addition, may remove some of the old links (decaying network) or, on the contrary, may add some new links between unconnected directly old sites (developing network) with the relative rate $c$. Therefore, to study even one-site properties of the structures, one has to keep in memory information about all connections among them. We performed simulations with a total number of sites (e. g. time) $t=1000$ with $100000$ averages for decaying networks, $-1<c< 0$, and $t=10000$ with $10000$ averages for developing networks, $c\geq0$. In Fig. 1, we present the dependences of the average connectivity $\overline{k}(s,t)$ on the number of a site $s$ at different values of $c$ for both structures, 
i.e. for the decaying network, Fig. 1 (a), and for the developing one, Fig. 1 (b). 

For both models, in the whole range of $c$, $\overline{k}(s,t) \propto (s/t)^{-\beta}$. The change of the sign of the exponent $\beta$ in the developing network at $c=-1/2$ was unexpected (compare with the behaviour of 
$\beta$ vs. an aging exponent in networks with aging of sites \cite{dm00}), see Fig. 1 (a) and the dependence $\beta(c)$ in Fig. 2. 
At this point, the average connectivity turns to be independent of the site age, $\overline{k}(s,t)=1$.
We studied also the distribution $P(k)$. 
It behaves as $k^{-\gamma}$ for all $c\geq 0$ for the developing network but, for the decaying network, the power-law dependence is found only in a narrow region of $c$ near zero (see Fig. 2). 
The range of the values of $k$ for which we observe such behaviour diminishes with decrease 
of $c$ and then disappears. Note that the finite size effects are strong in this region ($c<0$).      

We studied also the models in which, unlike the structures considered above, 
old links between sites are 
permanently removed with probability proportional to the product of the connectivities of the sites, or new links between old sites are permanently added with equal probability. 
The simulation demonstrates that the scaling breaks in both cases. 

Let us describe the obtained results analytically. 
We start from the case of the developing network.
One may use the simple continuous approach \cite{baj99,dm00} that gives exact results for the scaling exponents as it was demonstrated in \cite{dms00}. 
Since one site is added per unit of time, then the total number of sites is $t$ and each site is labeled by the time of its birth $s \leq t$. 
Then the equation for the average connectivity of the site $s$ at time $t$, $\overline{k}(s,t)$, in the continuous limit may be written in the following form:
\begin{equation}
\label{q1}
\frac{\partial \overline{k}(s,t)}{\partial t} = 
\frac{\overline{k}(s,t)}{\int\limits_0^t du \overline{k}(u,t)} + 
2c\, \frac{\overline{k}(s,t)\left[\int\limits_0^t du \overline{k}(u,t) - \overline{k}(s,t)\right]}
{\left[\int\limits_0^t du \overline{k}(u,t) \right]^2 - \int\limits_0^t du \overline{k}^2 (u,t)} 
\, , \ \ \ \ \ \ \ \ \overline{k}(t,t)  =  1 
\end{equation}
(we wrote also the boundary condition -- one link connected with a new site is added at each time step). The first term is the same as in the BA model, and the second one describes the increase of the connectivity due to the addition of new links between old sites with probability proportional to the product of connectivities of the connected sites. 
Note that the new links between the old sites can appear only if there is still no links between them. 
Hence, to write the last term in the present form, we make a strong assumption: we assume that the effect of multiplying of links is not essential at long times. We checked the validity of this non obvious assumption by simulation.  

Eq. (\ref{q1}) is 
simplified at long times:
\begin{equation}
\label{q2}
\frac{\partial \overline{k}(s,t)}{\partial t} = 
(1+2c)\frac{\overline{k}(s,t)}{\int\limits_0^t du \overline{k}(u,t)} \, .
\end{equation}
Applying $\int_0^t ds$ to Eq. (\ref{q2}) one gets 

\begin{equation}
\label{q3}
\int_ 0^t ds\frac{\partial \overline{k}(s,t)}{\partial t} =
\frac{\partial }{\partial t} \int_ 0^t ds \overline{k}(s,t) - \overline{k}(t,t) = 1+2c \, ,
\end{equation}  
so we obtain the obvious relation $\int_ 0^t ds \overline{k}(s,t) = 2(1+c)t$.  
%
Now Eq. (\ref{q1}) is of the following simple form:
\begin{equation}
\label{q5}
\frac{\partial \overline{k}(s,t)}{\partial t} = \frac{1+2c}{2(1+c)} \frac{\overline{k}(s,t)}{t} \, .
\end{equation} 
It solution is $\overline{k}(s,t) = (s/t)^{-\beta}$ with the exponent 
\begin{equation}
\label{q7}
\beta = \frac{1+2c}{2(1+c)} \, .
\end{equation}
To obtain the exponent of the distribution of connectivities, $\gamma$, one uses the general relation between the scaling exponents of growing networks \cite{dm00,dms00}:
\begin{equation}
\label{q8}
\beta(\gamma - 1) = 1 \, 
\end{equation}
that was obtained on the assumption that both $\overline{k}(s,t)$ and $P(k)$ show scaling behaviour. 
From our simulation, we know that this condition is fulfilled for $c > 0$.
Therefore,
\begin{equation}
\label{q9}
\gamma = 2 + \frac{1}{1+2c} \, ,
\end{equation}
so we get both scaling exponents for the developing network.

Let us consider now the decaying network.
Again we apply the continuous approach.  
In fact, the removal of old links with equal probability seems to be equivalent to the decrease of the connectivities of old sites with probability proportional to their particular values, so Eq. (\ref{q2}) may also be applicable to this case. Nevertheless, one should account for the fact that only existing links may be removed. Therefore, we prefer to make the calculations more thoroughly.

One introduces the average number of links between the sites $s$ and $s^\prime$ at time $t$, $\overline{n}(s,s^\prime,t)$, where $0 \leq s\leq s^\prime \leq t$. 
The average connectivity may be expressed in terms of this quantity: 
\begin{equation}
\label{q9a}
\overline{k}(s,t) = \int\limits_0^s du\, \overline{n}(u,s,t) + 
\int\limits_s^t dw\, \overline{n}(s,w,t) \, .
\end{equation}

The set of equations for $\overline{n}(s,s^\prime,t)$ is
\begin{eqnarray}
\label{q10}
& &  \overline{n}(s,t,t) = \frac{\int\limits_0^s du\, \overline{n}(u,s,t) + 
\int\limits_s^t dw\, \overline{n}(s,w,t)}
{\int\limits_0^t ds\left[\int\limits_0^s du\, \overline{n}(u,s,t) + 
\int\limits_s^t dw\, \overline{n}(s,w,t)  \right]} \, , 
\nonumber 
\\[9pt]
& & \frac{\partial \overline{n}(s,s^\prime,t)}{\partial t} = 
c\, \frac{\overline{n}(s,s^\prime,t)}{\int\limits_0^t ds \int\limits_s^t ds^\prime\, \overline{n}(s,s^\prime,t) } \, .
\end{eqnarray}
(Note that $c$ is negative now!) 
The first equality of Eq. (\ref{q10}) describes the links added to the network together with new sites as in the BA model. We again set the number of links connected with each new site 
to be unit. 
Applying $\int_0^t ds$ to this equality we get
\begin{equation}
\label{q10a}
\int_0^{s^\prime} ds\, \overline{n}(s,s^\prime,s^\prime) = 1 \, .
\end{equation}
The second equality of Eq. (\ref{q10}) shows how $\overline{n}(s,s^\prime,t)$ changes due to the removing of links between the old sites. 
Application of $\int_0^t ds\int_s^t ds^\prime$ to this equality leads to the other obvious relation, 
$\int_0^t ds\int_s^t ds^\prime\, \overline{n}(s,s^\prime,t) = (1+c)t$. 

Let us search the solution of Eq. (\ref{q10}) in the scaling form
\begin{equation}
\label{q11}
\overline{n}(s,s^\prime,t) = \frac{1}{t}\, {\cal N}\left(\frac{s}{t}, \frac{s^\prime}{t} \right)
\, .
\end{equation}
Then
 \begin{eqnarray}
\label{q12}
& & {\cal N}(\xi,1) = \frac{\int\limits_0^\xi d\zeta\, {\cal N}(\zeta,\xi) + 
\int\limits_\xi^1 d\zeta^\prime\, {\cal N}(\xi,\zeta^\prime)}
{\int\limits_0^1 d\xi\left[\int\limits_0^\xi d\zeta\, {\cal N}(\zeta,\xi) + 
\int\limits_\xi^1 d\zeta^\prime\, {\cal N}(\xi,\zeta^\prime)  \right]} \, , 
\nonumber 
\\[10pt]
& & \left[ 1 - \xi\frac{\partial}{\partial \xi} 
- \xi^\prime\frac{\partial}{\partial \xi^\prime}   \right] {\cal N}(\xi,\xi^\prime) =
 c\, \frac{{\cal N}(\xi,\xi^\prime)}
{\int\limits_0^1 d\xi \int\limits_\xi^1 d\xi^\prime\, {\cal N}(\xi,\xi^\prime) }
\end{eqnarray}
and
\begin{equation}
\label{q12a}
\int_0^1 d\xi\, {\cal N}(\xi,1) = 1 \, .
\end{equation}

One sees that
\begin{equation}
\label{q13}
\int_0^1 d\xi\left[\int_0^\xi d\zeta\, {\cal N}(\zeta,\xi) + 
\int_\xi^1 d\zeta^\prime\, {\cal N}(\xi,\zeta^\prime)  \right] = 2(1+c)
\end{equation}
and 
\begin{equation}
\label{q14}
\int_0^1 d\xi \int_\xi^1 d\xi^\prime\, {\cal N}(\xi,\xi^\prime) = (1+c) \, .
\end{equation}

%

The solution of Eq. (\ref{q12}) may be found in the form:
\begin{equation}
\label{q16}
{\cal N}(\xi,\xi^\prime) = B\, \xi^{-a} \xi^{\prime\,-b} \, ,
\end{equation}
where $B,a,b$ are constants. Inserting Eq. (\ref{q16}) into Eq. (\ref{q12}) with account for Eqs. (\ref{q13}) and (\ref{q14}) we obtain the exponents $a=b=1-1/[2(1+c)]$. Eq. (\ref{q12a}) gives $B=1-a=1/[2(1+c)]$. 
Substitution of Eq. (\ref{q16}) with account for Eq. (\ref{q11}) into Eq. (\ref{q9a}) leads to the expression $\beta=a=1-1/[2(1+c)]$ that is exactly the same as in the previous case, see Eq. (\ref{q7}). 
Note that now it is possible to use the relation between the scaling exponents, Eq. (\ref{q8}), only in the region near $c=0$, since only in this region we observed the scaling behaviour of $P(k)$. For these values of $c$ we again get the old expression Eq. (\ref{q9}) for the exponent $\gamma$. 

In Fig. 2, we plot the analytically obtained dependences $\beta$ and $\gamma$ vs. $c$ together with the results obtained from the simulation. One may see that the correspondence between the simulation and the theory is quantitative.  
When $c$ changes from $-1$ to $0$, $\beta$ increases from $-\infty$ to $1/2$ passing zero at $c=-1/2$. 
Subsequent increase of $c$ to $\infty$ leads to growth of $\beta$ up to $1$ while $\gamma$ decreases 
from $3$ to $2$. 
 The particular case $c \to \infty$, $\gamma=2$, corresponds to the situation when the network evolves only due to the addition of new links by the defined above rules, that resembles the original small-world networks of Watts and Strogatz \cite{ws98}.  

Our results show that the permanent removing of links leads to a more essential change of the structure of a network than the addition of them. What is the reason for that? One may see that the decaying structure under consideration is, in fact, a changing set of disconnected clusters. Because of finite size effects, we failed to find the position of the percolation threshold that may be defined for networks \cite{nw99}. Nevertheless, we see that, at high enough rates of link removal, large clusters are certainly absent, and the appearing structures indeed have to demonstrate quite different properties than the networks with $c \geq 0$. We failed also to find any peculiarity in the distribution of clusters in the point of the scaling break, $c=-1/2$.    
   
In summary, we have introduced a new parallel component of the evolution of growing networks. In addition to new links connecting new sites and old ones, links between old sites may appear or break with the relative rate $c$. We have demonstrated that addition of this component to a scale-free network does not break the scaling behaviour in a wide range of the rate $c$. 

The following questions remain open. 
What is the meaning of the point $c=-1/2$ in which the exponent $\beta$ changes sign? 
Is there any peculiar value of negative $c$ below which the power-law behavior of $P(k)$ breaks?  


SND thanks PRAXIS XXI (Portugal) for a research grant PRAXIS XXI/BCC/16418/98. JFFM was partially supported by the projects PRAXIS/2/2.1/FIS/299/94. We also thank M.C. Marques for reading the manuscript and A.V. Goltsev, Yu.G. Pogorelov and A.N. Samukhin for many useful discussions.\\
$^{\ast}$      Electronic address: sdorogov@fc.up.pt\\
$^{\dagger}$   Electronic address: jfmendes@fc.up.pt

\begin{figure}
\epsfxsize=3in
\begin{center}
\leavevmode
\end{center}
\caption{
Average connectivity vs. number $s$ of the site for (a) decaying ($-1<c<0$) and (b) evolving ($c>0$) networks. 
One site per unit of time is added. Different curves correspond to different values of $c$.
The network size is $t=10000$ (a) and $t=1000$ (b). 
}
\label{f1}
\end{figure}

\begin{figure}
\epsfxsize=2.2in
\begin{center}
\leavevmode
\end{center}
\caption{
Exponents $\beta$ of the average connectivity and $\gamma$ of the distribution of connectivities vs. $c$,  
i.e. vs. the rate of removal ($c<0$) or addition ($c>0$) links between old sites. Points are obtained from the simulations. The lines are found analytically [see Eqs. (\protect\ref{q7}) and (\protect\ref{q9})]. For the decaying network, the scaling behaviour of $P(k)$ is observed only in narrow region of $c$ close to zero. 
}
\label{f2}
\end{figure}


\end{document}